\documentclass[aps,pra,reprint,superscriptaddress,nofootinbib,floatfix]{revtex4-2}
\usepackage[utf8]{inputenc}
\usepackage{amsmath}
\usepackage{amssymb}
\usepackage{physics}
\usepackage{bbold}
\usepackage{graphicx}
\usepackage{multirow}
\usepackage{hyperref}
\hypersetup{colorlinks=true, citecolor=blue}
\usepackage{cleveref}
\usepackage{xcolor}
\usepackage{xspace}
\usepackage{bbold} 
\usepackage{orcidlink}

\usepackage[caption=false,font=small]{subfig}

\newcommand{\cm}{\ensuremath{\,\text{cm}}\xspace}
\newcommand{\mm}{\ensuremath{\,\text{mm}}\xspace}
\newcommand{\nm}{\ensuremath{\,\text{nm}}\xspace}

\newcommand{\MeV}{\ensuremath{\,\text{Me\hspace{-.08em}V}}\xspace}
\newcommand{\GeV}{\ensuremath{\,\text{Ge\hspace{-.08em}V}}\xspace}

\newcommand{\ps}{\ensuremath{\,\text{ps}}\xspace}

\usepackage[splitrule]{footmisc}
\begin{document}

\title{Quantum Dot Based Chromatic Calorimetry: A proposal}

\author{Yacine Haddad\orcidlink{0000-0003-4916-7752}}
\email{yacine.haddad@cern.ch}
\affiliation{Northeastern University, Boston, Massachusetts, USA}

\author{Devanshi Arora}
\affiliation{Shizuoka University, Shizuoka 432-8011, Japan}

\author{Etiennette Auffray\orcidlink{https://orcid.org/0000-0001-8540-1097}}
\affiliation{CERN, European Organization for Nuclear Research, Geneva, Switzerland}

\author{Michael Doser\orcidlink{https://orcid.org/0000-0002-3618-0889}}
\affiliation{CERN, European Organization for Nuclear Research, Geneva, Switzerland}

\author{Matteo Salomoni}
\affiliation{University of Milano-Bicocca. Piazza dell'Ateneo Nuovo, 1, 20126 Milan, Italy}

\author{Michele Weber\orcidlink{https://orcid.org/0000-0002-2770-9031}}
\affiliation{University of Bern, CH-3012 Bern, Switzerland}

\begin{abstract}

Chromatic calorimetry introduces a novel approach to calorimeter design in High Energy Physics (HEP) by integrating Quantum Dot (QD) technology into traditional homogeneous calorimeters. The tunable emission spectra of QDs provide new possibilities for energy reconstruction and potentially improved particle identification in homogeneous devices. This paper presents the initial conceptual design of a chromatic calorimeter aimed at validating the feasibility of QD-based chromatic calorimetry. The study evaluates its expected performance through detailed simulations, comparing the results to the existing calorimetric technologies. By embedding QDs with distinct emission properties within scintillators, we aim to improve both timing resolution and longitudinal segmentation, thus opening new avenues for precision measurements in future HEP experiments. While further development and validation are required, QD-enhanced detectors may represent a viable option for future HEP experiments, providing an additional tool for addressing the evolving demands of particle detection.

\end{abstract}
\maketitle

\section{Introduction}
Calorimetry plays a critical role in high-energy physics, providing essential particle energy measurements and enabling event kinematics reconstruction. Traditional calorimeters, whether homogeneous or sampling-based, have reached impressive levels of precision. However, there remains a significant challenge in enhancing the resolution and efficiency, particularly when aiming to meet the demands of future HEP experiments~\cite{linssen_physics_2012}. 

To take full advantage of the experimental opportunities created by such colliders, new and innovative particle detectors will be needed, as the quality of the scientific output will depend on the quality of the detectors with which experiments at these machines will be performed. In these experiments, that quality primarily concerns the precision with which the four vectors of the scattered objects produced in the collisions can be reconstructed and measured. Among the objects targeted by these experiments are leptons, photons, quarks, and gluons. Quarks and gluons manifest as a ``jet'' of secondary particles. Achieving the best possible precision for the momentum and energy measurements of jets is one of the most essential design goals of the proposed experiments.

In particle physics and related fields, a calorimeter is a detector in which the particles to be detected are entirely absorbed~\cite{livan_calorimetry_2019}. The detectors provide a signal proportional to the energy deposited in the absorption process. They play a crucial role in jet reconstruction, neutrino detection and particle identification, especially when combined with other sub-detectors.

As we look towards the future of high-energy physics, particularly with projects like the Future Circular Collider~\cite{aime_muon_2022,abada_fcc_2019} (FCC), it becomes clear that existing technologies, though proficient for current tasks, are approaching their operational boundaries. The FCC and similar next-generation collider facilities, including both the hadron collider FCC-hh~\cite{benedikt_fcc-hh_2019} and the lepton collider FCC-ee~\cite{benedikt_fcc-ee_2019}, present distinct but equally demanding challenges for detector design, particularly in calorimetry. The FCC-hh faces unprecedented pileup rates and multiple collisions per bunch crossing, complicating event reconstruction, data interpretation, and the identification of rare decay modes or new particles\footnote{The High-Luminosity Large Hadron Collider (HL-LHC) is expected to increase luminosity by a factor of 5-7 times that of the original LHC design ($\rm 1\times 10^{34}~\cm^{-2}s^{-1}$), while the circular hadron collider FCC-hh aims to enhance this figure by an order of magnitude. The expected pile-up is around $140-200$ for HL-LHC\cite{apollinari_high-luminosity_2015} and could exceed 1000 for FCC-hh~\cite{abada_fcc_2019}.}. These challenges place stringent demands on calorimeter timing resolution and energy reconstruction, with requirements for advanced longitudinal segmentation to disentangle overlapping events and achieve precise jet energy measurements in a high-radiation environment. In contrast, future lepton collider projects such as the FCC-ee, CLIC~\cite{linssen_physics_2012}, and ILC~\cite{behnke_international_2013} emphasise precision over event rate, necessitating calorimeter designs capable of achieving unparalleled energy and angular resolutions. The physics program for the FCC-ee, for example, requires calorimeters optimised for minimal systematic uncertainties, such as achieving higher angular resolutions, precise jet energy measurements, and heavy-flavour tagging. Technologies under consideration include highly granular electromagnetic~\cite{noauthor_phase-2_2017} calorimeters and dual-readout systems~\cite{lee_dual-readout_2018}, which have demonstrated feasibility for meeting these stringent requirements.

The current state-of-the-art in homogeneous calorimetry, predominantly relying on monolithic crystals, faces critical limitations in two key areas: timing resolution and longitudinal segmentation. The typical timing resolution of about $200\ps$ per detection cell falls short of the demands of future collider projects. However, CMS~\cite{noauthor_phase-2_2017} at the HL-LHC targets a resolution between $30$ and $50~ps$, demonstrating that such precision is achievable within this timescale. Highlighting these advancements underscores the importance of enhanced timing precision to accurately track and differentiate between the multitude of particle events occurring in rapid succession.

The limitation in longitudinal segmentation is equally significant. Longitudinal segmentation refers to the ability to distinguish different layers within a particle shower, providing detailed information on the development and composition of the shower. This segmentation is crucial for understanding the complex interactions of particles, especially in high-energy collisions, as it allows for more precise reconstruction of events and better identification of particle types~\cite{brient_improving_2005}. Accurate particle detection is not only essential for identifying and analysing particle interactions precisely but also for isolating specific events of interest from a dense and complex background. Current calorimetry methods and technologies, such as the homogeneous electromagnetic calorimeter in the CMS experiment~\cite{noauthor_cms_nodate}, lack the capability for effective longitudinal segmentation, hindering the depth of analysis possible in high-energy physics experiments.

This paper introduces a new calorimetry concept that utilises the properties of QD technology to address the challenges of longitudinal segmentation in homogeneous electromagnetic calorimeters. By integrating QDs into scintillators, this concept aspires to introduce a new approach to longitudinal segmentation. This methodology, termed as Chromatic Calorimetry~\cite{doser_quantum_2022} (CCAL), takes advantage of the tuneability and narrow emission bandwidth ($\sim 20\nm$) of quantum dots, quantum wells, carbonised polymer dots, monolayer assemblies, and nanocrystals\cite{liu_deep_2020,zhang_distributed_2021,yuan_engineering_2018}. By embedding QDs with varying emission wavelengths into a single transparent, high-density material, the CCAL aims to create a calorimeter module that differentiates the shower stages by colour. QDs emitting longer wavelengths would be placed at the front, and those with shorter wavelengths at the back. With the demonstrated $20\nm$ emission bandwidths and confining emissions to the visible spectrum, it is possible to distinguish up to twenty unique emission zones, allowing for precise measurement of the shower's evolution. This feature enables accurate mapping of radiation's position and intensity within the module to the wavelength and intensity of the produced fluorescence light. 

The current paper serves as a proposal for the concept of chromatic calorimetry and relies on Geant4 simulations as the primary tool for validating key design parameters. These simulations complement the ongoing experimental work by providing a robust proof-of-concept and demonstrating the feasibility of this innovative technology. Specifically, we simulate the effects of embedding QDs with distinct emission properties into transparent polymers or plastic scintillators. While experimental results are beyond the scope of this paper, initial efforts using conventional scintillating materials have already begun to validate the concept of chromatic reconstruction~\cite{salomoni_enhancing_2024,arora_enhancing_2024}. This proposal lays the foundation for future experimental studies necessary to fine-tune the design and achieve full validation.

\section{Background and Motivation}
The emergence of quantum technologies has a strong potential to enhance detection capabilities and explore phenomena at multiple energy scales, which represents a promising avenue for future collider experiments. In the last few years, the HEP community has started expressing interest in exploring these new detection techniques. The Snowmass report~\cite{humble_snowmass_2022,alam_quantum_2022} as well as the ECFA report~\cite{ecfa_detector_rd_roadmap_process_group_2021_2020} have both suggested promising directions that the community could follow in the next few years. Among these, one of the most mature technologies for particle detection is nanocrystals. These semiconductor nanocrystals possess quantum mechanical properties that confer upon them the ability to emit light at specific frequencies when excited.

\begin{figure*}[!ht]
    \subfloat[][]{\includegraphics[width=.3\textwidth]{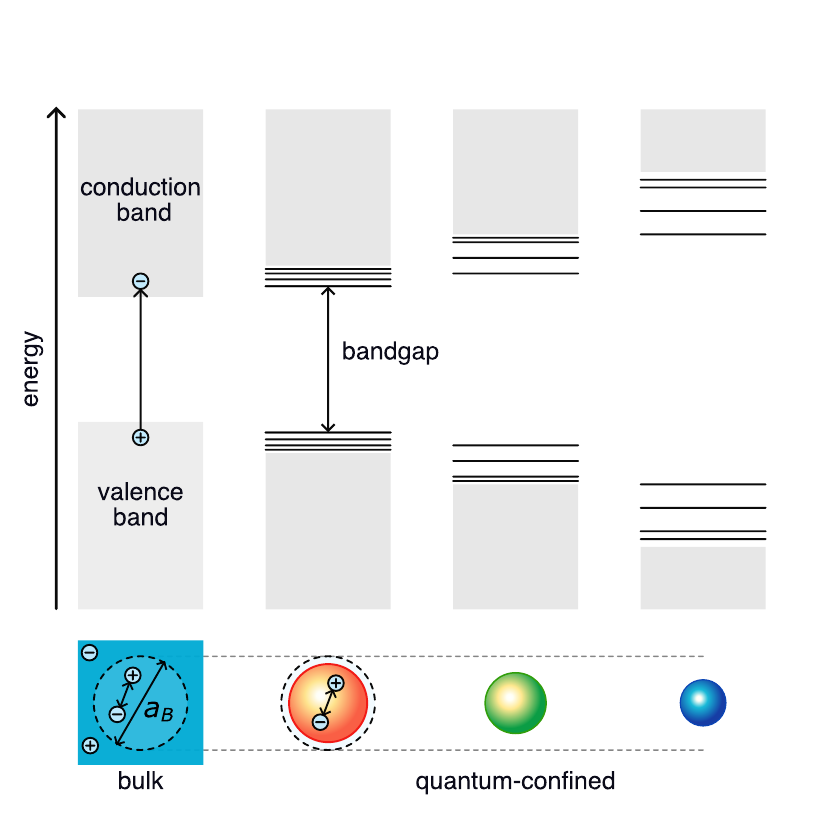}\label{fig:confinement-qd}}%
    \hspace{2cm}
    \subfloat[][]{\includegraphics[width=.42\textwidth]{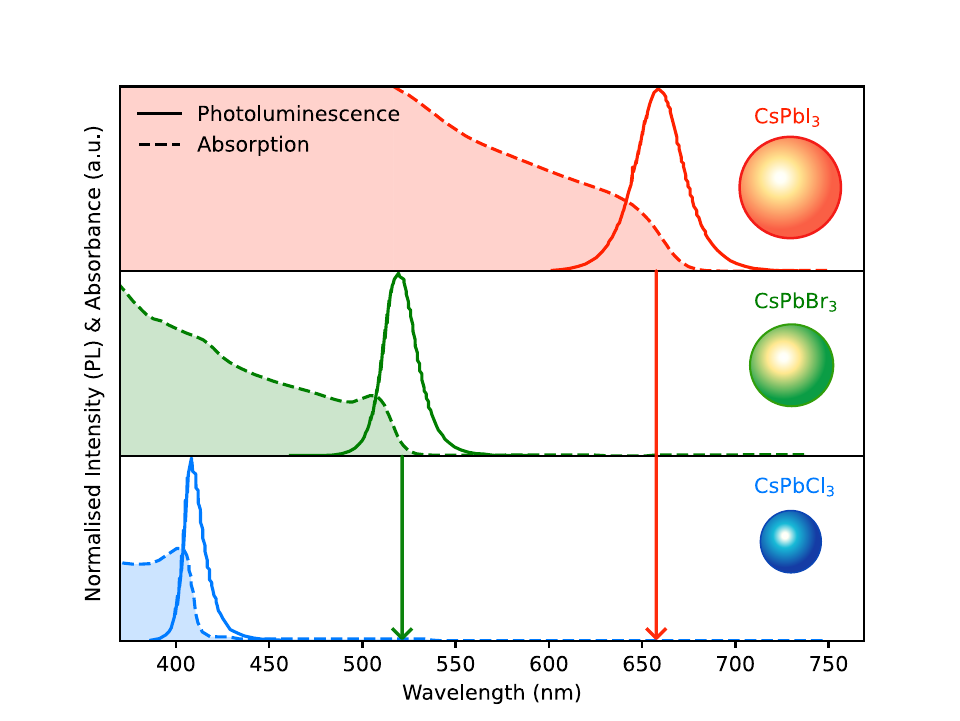}\label{fig:specra-qd-light}}%
    \caption{
        (\ref{fig:confinement-qd}) Quantum confinement, which leads to size-dependent optical and electrical properties distinct from those of the parental bulk solid, occurs when the spatial extent of the electronic wave functions is smaller than the Bohr exciton diameter $a_b$. (\ref{fig:specra-qd-light}) The graph illustrates the absorption (dashed lines) and photoluminescence (PL) emission (solid lines) spectra for QDs of different sizes (data from~\cite{zheng_near-infrared-triggered_2018}). Smaller CsPbCl3 QDs, with blue photoluminescence, do not absorb the red photoluminescence from the larger CsPbI3 QDs due to their absorption spectra being at shorter wavelengths. This phenomenon enables non-interfering emissions across different wavelengths, determined by size, which can be utilised to engineer a transparent medium with directionally layered light emission. 
    }
    \label{fig:diff-qd-emission}
\end{figure*}

Semiconductor nanocrystals are classified into three main categories: Quantum wells, wires, and dots. The available energy levels in such objects are discretised as a function of the object dimensionality (0D, 1D and 2D, respectively) and their size and shape~\cite{cassidy_nanoshell_2020,chen_all-inorganic_2018,roda_understanding_2021,meng_perspectives_2021}. Quantum confinement emerges when electrons are constrained to a domain comparable with their de Broglie wavelength. In quantum dots, electrons and holes exhibit a discrete, or quantised, atomic-like density of states (see Figure.~\ref{fig:confinement-qd}). As the QD size decreases, quantum confinement increases the effective bandgap, leading to a blue shift in the absorption and emission spectra. It should be stressed, however, that size is only one of several tuning knobs: the bandgap may also be engineered through chemical composition, for instance by substituting different halide ions (Cl, Br, or I) in all-inorganic perovskites of the form CsPbX, where X represents the halogen element. In the latter case, illustrated schematically in Figure.~\ref{fig:diff-qd-emission}, the halogen atom is the principal driver of the emission energy, while size plays a secondary role at the few-nanometre scale \cite{williams_perovskite_2020,chen_all-inorganic_2018}
An example of a gradually stepping emission is shown in Figure.~\ref{fig:specra-qd-light}. QDs have already found many applications in the industry, from improving solar panels and screens to medical imaging~\cite{chen_all-inorganic_2018,krumer_tackling_2013}.

When integrated with conventional scintillator materials\footnote{Throughout this paper, the term \emph{conventional scintillator} designates bulk inorganic or organic materials whose scintillation light originates from intrinsic lattice states or dopant activators. These traditional scintillators do not incorporate nanostructured emitters such as quantum dots or nanocrystals. For examples LYSO:Ce (HGCAL\cite{noauthor_phase-2_2017}) and the plastic scintillators such as EJ-200 (ATLAS TileCal\cite{atlas_collaboration_atlas_1997}) are both conventional scintillator.} or a supporting polymeric matrix, semiconductor nanocrystals from heterostructures or Meta-Crystals~\cite{lecoq_metamaterials_2008,noauthor_tical_2014,turtos_use_2019}, these hybrid materials combine the high absorption efficiency of traditional scintillators with the rapid light emission properties of nanocrystals, thereby enhancing both energy and time resolution~\cite{mccall_fast_2020,carulli_stokes_2022}. Conventional scintillators produce light proportional to the energy deposited by charged or neutral particles. The energy transfer from initial ionisation in the bulk material to the luminescence centres is complex. It leads to an intrinsic time-resolution limit in photoproduction due to the stochastic relaxation processes of the hot electron-hole pairs produced by the impact of radiation on the crystal material. Overcoming these limitations requires faster photon generation mechanisms, and quantum dot-based materials offer a promising alternative~\cite{li_are_2023}. The unique properties of quantum dots, including their tunable emission frequencies and rapid response times, make them an attractive candidate for advancing timing precision in particle detection. Current research efforts, particularly in particle physics and medical imaging, are exploring the potential of quantum dots to complement traditional scintillators, with preliminary findings supporting their ability to significantly improve performance~\cite{williams_perovskite_2020,carulli_stokes_2022,mccall_fast_2020,vanecek_advanced_2022}.

\section{Designing a Chromatic Calorimeter}
\begin{figure}[!ht]
    \centering
    \includegraphics[width=0.4\textwidth]{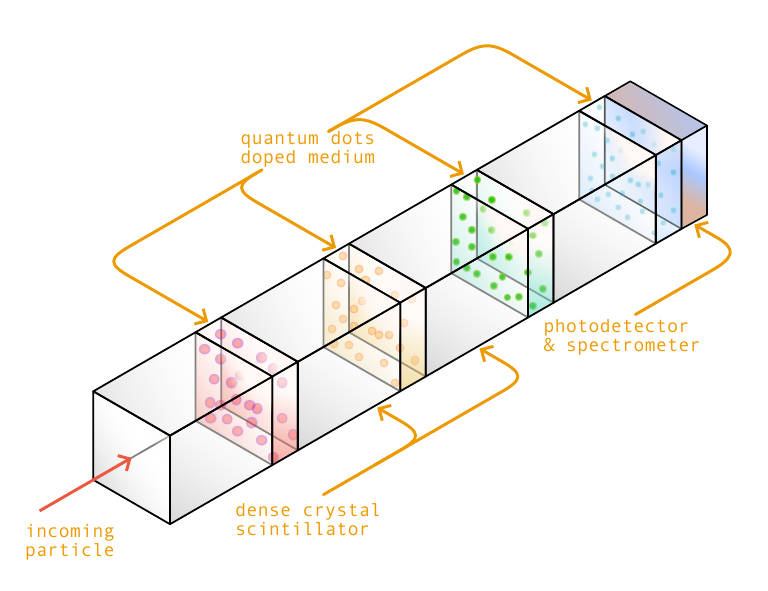}
    \caption{\small  Diagram of Chromatic Calorimeter Module employing standard dense crystal scintillators with interlayered doped organic scintillators serving as wavelength converters. It features a photodetector and a spectrometer at the module's rear to measure the scintillation light produced by the different crystal layers.}
    \label{fig:ccal-sampled}
\end{figure}

The properties described in the previous section create an opportunity for a new method of measuring electromagnetic or hadronic showers within a scintillator. In this context, the concept proposed here enables a detailed view of the shower profile via a chromatic reconstruction. In such a device, a CCAL module must be constructed from multiple high-density transparent materials. Each material would be doped along its length with QDs of different emission wavelengths, arranged from longest to shortest (see Figure.~\ref{fig:ccal-sampled}). The arrangement from the larger wavelength emitting QD at the entrance of the incoming particle to the shortest allows successive transparency, where the red light propagates all the way down to the photodetector without being absorbed by the deeper layers, as shown in Figure.~\ref{fig:specra-qd-light}. With current technology allowing for 20\nm emission bandwidths, around twenty different emission regions are conceivable, facilitating fine-grained measurements of shower development. However, the radiation tolerance of these nanocrystals must be determined.

A CCAL module is a polychromatic-embedded wavelength shifter, mapping radiation's position and intensity within the module to the wavelength and intensity of the produced fluorescence light. Multiple emission regions could be uniquely identified in a single measurement. Challenges include incorporating QDs into existing dense crystals with a high atomic number (high-Z) during growth or alternatively embedding QDs within a low-density polymer material.

There are two possible approaches to designing a chromatic calorimeter to take advantage of QD properties. The first is what we call the \textbf{direct embedding} approach. This approach relies on integrating QDs like caesium lead halide perovskite nanocrystals (CsPbX3) into high-Z inorganic scintillators such as Bismuth Germanate (BGO) and Lead Tungstate (PWO). Inspired by recent advancements in embedding QDs into glass ceramics~\cite{ma_highly_2021,su_enhanced_2022,tong_enhanced_2021}, this novel method requires balancing the melting points of QDs and scintillator materials to maintain the QDs' structural integrity and luminescent properties, thereby tuning the scintillation properties and spectral resolution~\cite{su_enhanced_2022}. Exploring the potential of embedding QDs into high-Z transparent materials can also be considered, assuming that the QDs will impart scintillation properties to the host medium\footnote{Recent studies have explored embedding QDs into transparent matrices to enhance scintillation performance, such as in polymer matrices~\cite{liu_transparent_2017}, borosilicate glass~\cite{su_enhanced_2022}, and Glass-Ceramic~\cite{ma_highly_2021}.}.

A speculative yet potentially groundbreaking method within this approach is the mixing of QD powder\footnote{An example of such powder could be from QDot$\rm ^{TM}$~\cite{noauthor_qdot_nodate}. This commercial powder is a composite material containing Perovskite CsPbBr3 QDs that have been embedded into a transparent Cs4PbBr6 matrix with the raw materials of the scintillating crystal before the melting phase~\cite{noauthor_qdot_nodate}}. This method, demanding a thoroughly homogenised mixture for even QD distribution, will proceed through the high-temperature Czochralski process~\cite{taranyuk_state_2019}, followed by quality control tests to confirm QD incorporation and assess modified scintillating properties. 

The second approach is what we call the \textbf{hybrid approach} shown in Figure.~\ref{fig:ccal-sampled}, which takes advantage of the recent research on QDs embedding in organic polymers or scintillators \cite{perego_highly_2022,perego_composite_2021,gramuglia_light_2021,vanecek_advanced_2022,turtos_ultrafast_2016,carulli_stokes_2022,williams_perovskite_2020}. In this approach, the quantum dots can be embedded in a polymer matrix such as Polymethyl Methacrylate (PMMA) and then interlayered with inorganic crystals handling the shower initiation and containment. This approach requires determining key parameters such as quantum dot concentration, transparency, radiation hardness, time response, and light yield for different quantum dot and matrix combinations. Considering the potential limitations related to the melting points and thermal stability of QDs when directly embedding them into high-Z scintillating crystals, the hybrid QDs approach emerges as an alternative worth consideration. 
 
To interpret the shower profile from the spectral intensity distribution, the photodetector must be capable of discerning the intensities and timings of individual spectral lines. The subsequent phase involves capturing and measuring light, given that the detection principle is based on capturing the yielded scintillation and Cherenkov lights while simultaneously measuring their wavelengths. This can be achieved by using photodetectors coupled with optical band-pass filters carefully selected to match the emission spectra of the QDs. The design relies on the transparency of the bulk material, with minimised self-absorption, to allow the light from different layers to reach the photodetectors at the rear of the calorimeter module.

Such an approach is experimentally possible with current technology. Photon counting can be done using the standard Silicon Photomultipliers (SiPMs), which are now widely available on the market. In contrast, the wavelength measurement should be done in a separate spectrometer device. Various micro-spectrometers are now available, such as the ultra-compact \texttt{C12666MA} spectrometer from Hamamatsu or the multispectral \texttt{AS7262} sensor-on-chip solution from ams-OSRAM. For example, the \texttt{AS7262}, a multispectral sensor-on-chip, could be an efficient and cost-effective spectral reconstruction solution. This sensor has six-channel spectral sensing across visible wavelengths (approximately 430~\nm to 670~\nm) with a full-width half-max (FWHM) of 40~\nm. Selecting QDs tuned to this chip's detection peaks would make it a compelling choice for precise spectral identification in the CCAL calorimeter.

Alternatively, more recent compact spectrometers based on nanowires~\cite{li_-chip_2021} or nanodots~\cite{miao_recent_2019,zhou_leadfree_2019} have been developed or are in active development. More traditional, though bulkier, alternatives like Bragg spectrometers or prismatic structures coupled to photodiode arrays are also conceivable. These options are worth exploring and should be considered should they become available.

Given the experimental difficulties of embedding QDs into high-Z materials, this paper will focus on the hybrid approach. The aim is to assess the workability of the chromatic reconstructions technique. We are currently exploring experimentally this idea with very promising preliminary results~\cite{arora_enhancing_2024,salomoni_enhancing_2024}.

\section{\textit{Ab Initio} Simulation}
To evaluate the performance of the proposed hybrid chromatic calorimeter, we conducted detailed simulations using the Geant4 toolkit~\cite{agostinelli_geant4simulation_2003,allison_geant4_2006}. Geant4 is a widely used platform for simulating the passage of particles through matter, particularly in HEP. It should be noted that Geant4 currently lacks built-in models for simulating complex interactions specific to QDs or nanostructures. Given this limitation, we employed a simplified modelling approach to approximate the optical behaviour of QDs while acknowledging that future experimental work will be necessary to fully validate these approximations.

\begin{figure*}[!ht]
    \includegraphics[width=\linewidth]{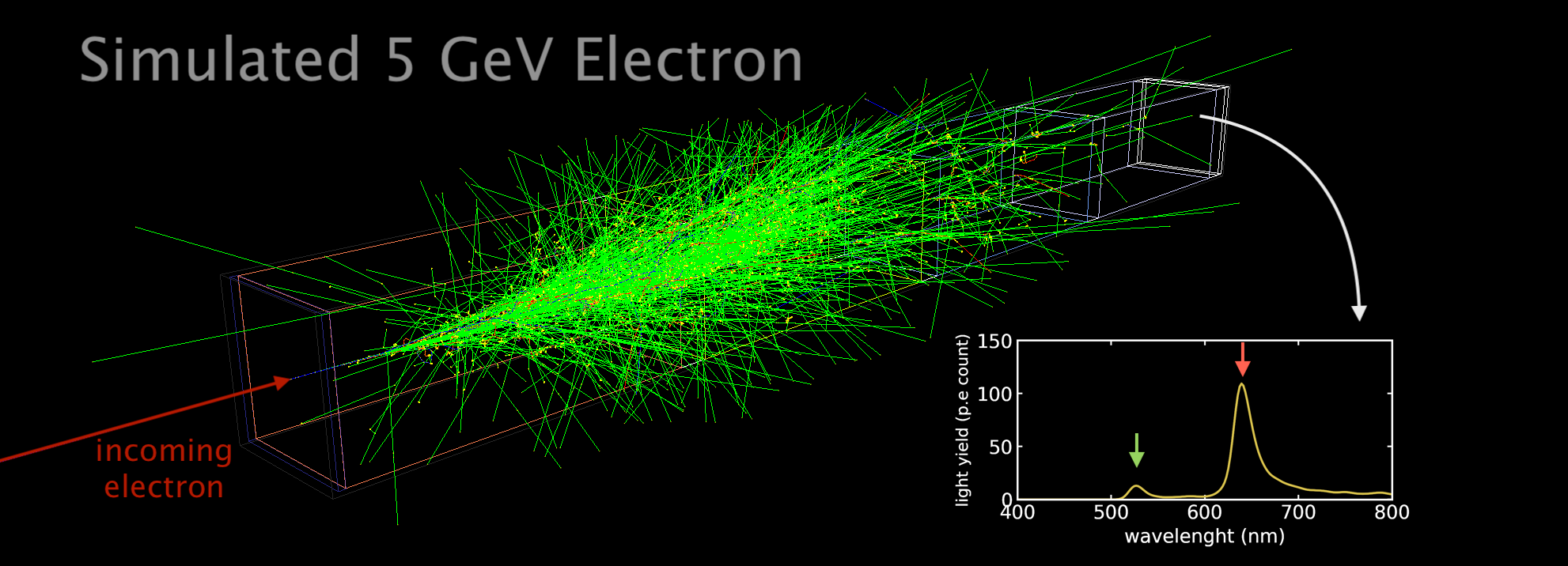}
    \caption{Simulation of a 5 GeV electron interacting with a CCAL module, consisting of four quantum dot-doped inorganic high-Z crystals and a photodetector positioned at the rear. The figure illustrates the electron-induced shower without depicting the scintillation light. Light propagates to the rear of the module, where photodetector channels, tuned to specific spectral peaks, capture the photons. These spectral peaks provide longitudinal information about the electromagnetic shower profile.}
    \label{fig:ccal-g4-event-display}
\end{figure*}

\begin{table}[!ht]
\centering
\caption{
Material parameters are taken from Ref.~\cite{particle_data_group_review_2022}.
}
\begin{itemize}\small
\item[$\dagger$] Refractive index at the emission maximum.
\item[$\star$] The emissions from PMMA are tunable by incorporating different quantum dots.
\end{itemize}

\label{tab:properties-mat}
\begin{tabular}{lcccccc} 
\hline
 & $\rho$   & $X_0$ & $\lambda_I$ & $n^\dagger$ & $\lambda_\text{max}$ & Light Yield  \\ 
 & $\rm g/cm^3$ & $\cm$  & $\cm$        &             & $\nm$            &              \\ \hline

PMMA     & $1.19$   & $34.07$ & $69.54$      & $1.49$ & $\star$  & -            \\ 
PWO & $8.30$   &  $0.89$ & $20.7$       & $2.20$ & 425      & $200~\text{photons}/\MeV$\\ 
\hline
\end{tabular}
\end{table}

We focused on simulating the hybrid approach of the chromatic calorimeter, which is considered feasible with current technology and can be implemented in the near future. The calorimeter design consists of four blocks of PWO crystals, each measuring $20\times20\times60~\mm^3$ equivalent to $\approx 26X_0$, acting as the primary scintillation medium. These PWO blocks are interleaved with $5~\mm$ thick layers of PMMA doped with QDs, serving as wavelength-shifting materials. This arrangement is depicted schematically in Figure.~\ref{fig:ccal-sampled}. Finally, in order to contain the scintillation light, the stack in wrapped in $0.5\mm$ reflective aluminium alveola. All the elements used in the simulation are modelled with full optical‐surface properties enabled.

The PWO crystals and the QD-PMMA sheets are pressed together in direct optical contact. No silicone grease or adhesive is used in the simulation\footnote{In a real setting an optical‐coupling layer (e.g.\ radiation-hard silicone grease or epoxy, $n \simeq 1.46–1.49$) would be applied to fill micro-gaps and ensure mechanical stability. Such a layer experimentally lowers the per-interface Fresnel loss to $\sim2–3\%$, so the simulation presented here is conservative.}. Each boundary is implemented in Geant4 as a polished dielectric-dielectric surface; thus, Fresnel reflection and refraction are treated natively. With refractive indices reported in Table.~\ref{tab:properties-mat}, the angular–averaged reflectance is $\langle R\rangle\approx 8\%$ per interface, giving a cumulative transmission $T=(1-\langle R\rangle)^2 \approx 85\%$. Since the stack is housed inside a $90\%$ specular aluminium alveola, photons that escape laterally are largely redirected inward, so the net unrecoverable loss is less than $\sim2\%$. Bulk Rayleigh scattering in both PWO and PMMA is also enabled; its contribution is $<1 \%$ photon loss per $\cm$, which is negligible compared with interface reflections.

At the rear of the module, we place four Silicon Photomultiplier (SiPM) detectors, each equipped with a band-pass optical filter tuned to the peak emission wavelength of the corresponding QDs used in the PMMA layers. For this simulation, we used the parameters from the Hamamatsu S14160~\cite{noauthor_mppc_nodate} to achieve higher Photon Detection Efficiency (PDE) peaking at 51\%. The filters are chosen from available band-pass filters provided by Thorlabs~\cite{noauthor_hard-coated_nodate}. However, these filters may have limited angular acceptance, which could impact performance. To address this, it is essential to consider filters with wider angular acceptance during the detector conception to ensure accurate measurements across various incident angles. Alternative options, such as the Wratten 2 filters from Kodak~\cite{noauthor_kodak_nodate}, could also be explored for their suitability in this application. Table~\ref{tab:qd-filters} summarises the emission peaks of the QDs and optical filter specifications used in the simulation.

\begin{table}[ht]
\centering
\caption{Emission peaks of the QDs and corresponding optical filters used in the simulation.}

\label{tab:qd-filters}
\begin{tabular}{lccccc}
\hline
Label & $\lambda^{\text{QD}}_{\rm max}$ & Filter & PDE & $\sigma_{\text{abs}}(400\text{nm})$ & $\ell_{\text{abs}}(400\text{nm})$\\
& nm & nm &  & $10^{-14}~\rm cm^2$ & cm \\
\hline
Red    & $630$ & $630\pm30$ & 25\% & 31.69&  0.03\\
Green  & $519$ & $520\pm30$ & 42\% & 14.79&  0.07\\
Blue   & $463$ & $450\pm30$ & 51\% &  6.94&  0.14\\
Purple & $407$ & $400\pm30$ & 47\% &  5.78&  0.17\\
\hline
\end{tabular}
\end{table}

In Geant4, we enable the \texttt{FTFP\_BERT} physics list for hadronic processes. We also enable the \texttt{EmStandardPhysics} for electromagnetic interactions and augment it with the \texttt{OpticalPhysics} module to track scintillation, Cherenkov, absorption, re-emission, and boundary reflections.  To comprehensively model particle interactions, our Geant4 setup employs the \texttt{FTFP\_BERT} physics list for hadronic processes and \texttt{EmStandardPhysics} for electromagnetic interactions~\cite{ivanchenko_recent_2011}. Furthermore, the \texttt{OpticalPhysics} module is included to account for detailed optical phenomena such as scintillation, Cherenkov radiation, absorption, re-emission, and boundary reflections.

\subsection*{Simulating QD response}

To model the absorption and re-emission processes in the QD-doped PMMA layers within Geant4, we aim to estimate the mean free path ($\ell$) of photons in a polymer matrix, specifically PMMA containing embedded QDs. The mean free path defines the average distance a photon travels before being absorbed by a QD, and this parameter is critical for simulating light propagation, absorption, and emission within the quantum dot-doped polymeric layer.

Since we rely on Geant4 to emulate the QD response, we use the G4OpWLS (Optical Wavelength Shifting) process, which models the wavelength-shifting behaviour of materials like those doped with QDs. The G4OpWLS process requires the interaction mean free path as input to simulate photon absorption and generate optical photons with a targeted QD emission peak. By providing the estimated mean free path and the emission properties of the QDs, we can effectively model the absorption and re-emission events in the QD-doped PMMA layers within Geant4.

The absorption cross-section ($\sigma_{\text{abs}}(\lambda)$) quantifies how much light is absorbed by a single quantum dot embedded in the PMMA matrix. This value is typically determined using UV-Vis absorption spectroscopy or sourced from literature. The absorption cross-section varies depending on the quantum dot size, material composition, and the incident wavelength. For caesium lead halide perovskites (CsPbX$_3$, where X = Cl, Br, I), typical absorption cross-section values range from $10^{-15}$ to $10^{-14} \cm^2$, with the exact value depending on the synthesis conditions and quantum dot size \cite{leatherdale_absorption_2002, yu_absorption_2005, chen_size-_2017, zheng_near-infrared-triggered_2018}.

When quantum dots are embedded in a PMMA matrix, their concentration (C) is usually expressed as the number of quantum dots per unit volume of the polymer. This concentration can be adjusted by controlling the weight percentage of quantum dots mixed into the polymer during the fabrication process. For practical calculations, the number of quantum dots per cubic centimetre of PMMA can be estimated from the initial weight percentage, the density of PMMA, and the quantum dot density. The concentration $n_{\text{QD}}$ in units of particles per cubic meter can be written as:
\begin{equation}
   n_{\text{QD}} = C_{\text{f}} \times \frac{\rho_{\text{PMMA}}}{m_{\text{QD}}} 
\end{equation}

Where $C_{\text{f}}$ is the weight fraction of QDs in PMMA, $\rho_{\text{PMMA}}$ is the density of PMMA ($1.19~\rm g/\cm^3$), and $m_{\text{QD}}$ is the mass of a single quantum dot, calculated based on its composition and size.

The mean free path $\ell$ is calculated based on the absorption cross-section and the concentration of quantum dots embedded in the PMMA matrix. Using the number density of quantum dots $n_{\text{QD}}$ and the absorption cross-section $\sigma_{\text{abs}}$, the mean free path can be derived using the following equation:

\begin{equation}
  \ell_{\rm abs}(\lambda) = \frac{1}{n_{\text{QD}} \cdot \sigma_{\text{abs}}(\lambda)}  
\end{equation}

This equation gives the average distance a photon will travel within the PMMA matrix before encountering and being absorbed by a quantum dot.

Compared to liquid solutions, the distribution and interaction of quantum dots in a polymer matrix can differ significantly. The quantum dots may experience different environmental effects, such as altered local dielectric constants and changes in inter-dot spacing, which can influence their optical properties, including absorption cross-sections and emission characteristics. Additionally, polymer chains may partially quench the photoluminescence or affect the quantum dot surface passivation, slightly altering their absorption properties \cite{su_enhanced_2022}. For the current simulation, we neglect these effects, but we acknowledge that these should be modelled carefully for an accurate account of the QD response in embedded materials. 

By embedding quantum dots in a PMMA matrix, the mean free path of photons can be effectively controlled by tuning the quantum dot concentration $C_f$ and size. Hence, the absorption length for optical photons can then be set in Geant4 simulation properties. The concentration $n_{QD}$ for the current simulation is chosen to be $\sim 10^{14}~\cm^{-3}$, based on the prediction from \cite{yu_absorption_2005,moreels_size-dependent_2009}. This particle density corresponds to a molar concentration of $\sim 0.17~\mathrm{\mu M}$, a value selected to ensure a high probability of photon capture while remaining in a regime where severe self-absorption effects are not expected to dominate. While this serves as a robust starting point for simulation, we acknowledge that the ideal concentration for a physical device will require empirical validation to optimise the balance between signal intensity and quenching effects. A summary of the values used in the simulation is shown in Table ~\ref{tab:qd-filters}. 

\subsection{Photon Detection}
Using the calculated optical properties, we defined custom materials in Geant4 to represent the QD-doped PMMA layers. The absorption lengths and emission spectra were input into the simulation to model the re-emission of photons by the QDs. The emission spectra of the QDs were taken from the manufacturer’s data~\cite{zheng_near-infrared-triggered_2018} and were assumed to have a Gaussian distribution centred at the specified peak wavelengths. The scintillation properties of the PWO crystals were defined based on standard values available in the literature~\cite{particle_data_group_review_2022}. The scintillation yield, emission spectrum, and decay times were specified to accurately model the photon production in the PWO blocks.

We simulated the passage of electrons through the calorimeter to induce electromagnetic showers. The Cherenkov and scintillation processes within the PWO crystals were modelled using Geant4’s built-in physics processes~\cite{janecek_simulating_2010}. Optical photons generated in the PWO blocks are absorbed and re-emitted by the QD-doped PMMA layers according to the defined absorption and emission properties.

The SiPM detectors at the rear of the calorimeter were modelled as ideal photon counters, with their spectral response shaped by the band-pass filters specified in Table~\ref{tab:qd-filters}. This allowed us to record the number of photons detected in each wavelength band corresponding to the different QD emissions. We conducted simulations with electron beams of various energies ranging from $5~\GeV$ to $100~\GeV$ to study the calorimeter response as a function of incident particle energy. Each simulation run involved tracking a statistically significant number of events (typically $10^4$) to ensure reliable results. Each channel's response is estimated by counting the number of photons.

\subsection{MIP calibration}
\label{MIP-Calibration}
Finally, we simulated the interaction of minimum ionising particles (MIPs), specifically muons at different energies, to calibrate the inter-channel response of the calorimeter. Since MIPs should, in principle, deposit the same energy in each crystal layer, the readout should reflect this uniformity. We define an inter-channel correction factor as the expected ratio of the amplitude of the reference channel (taken here as the response of the first channel) to the amplitude of each channel:

\begin{equation}
    c_i(E_\text{beam}) =  \biggl<\frac{A_0(E_\text{beam})}{A_i(E_\text{beam})}\biggl>_\text{mip}
\end{equation}

Where $A_i(E_\text{beam})$ is the amplitude of the $i-$th channel at beam energy $E_\text{beam}$, and  $\langle\rangle_\text{mip}$ denotes the average value over MIP events\footnote{The dependence of the correction factor on beam energy arises because we expect a change in amplitude due to the increase in the number of photons produced by Cherenkov radiation. The number of Cherenkov photons has a dependence that follows  $N_c \propto (1 - 1/\beta n(\lambda))$, where $\beta$ is the particle’s relative velocity and  $n(\lambda)$ is the wavelength-dependent refractive index}. This correction factor allows us to rescale each channel with the same energy response. The response of each layer is then defined as:

\begin{equation}
 R_i(E_\text{beam}) = c_i(E_\text{beam}) A_i(E_\text{beam})
\end{equation}

For MIPs, this ensures that the average response at each channel equals that of the reference channel, i.e.,  $\langle R_i \rangle = \langle R_0 \rangle$. To express the response in units of MIPs, we normalise the corrected response at each beam energy by the corrected MIP response of the reference channel $\langle R_0 \rangle_{\text{mip}}$. The final corrected response is given by:

\begin{equation}
 R_i^c(E_{b})[{\rm MIP}] = \frac{R_i(E_{b})}{\langle R_0(E_b)\rangle_\text{mip}} = c_i(E_{b})  \frac{A_i(E_{b})}{\langle R_0(E_b)\rangle_\text{mip}}
\end{equation}
The correction factor encapsulates various effects that must be studied separately in future iterations. For instance, each channel sits at a different photodetector PDE. Various reflections occur at each layer, and these reflections are the same at each layer. There are also possible attenuations that need to be studied and evaluated experimentally.

\section{Simulation Results \& Discussion}

\begin{figure}
    \includegraphics[width=0.5\textwidth]{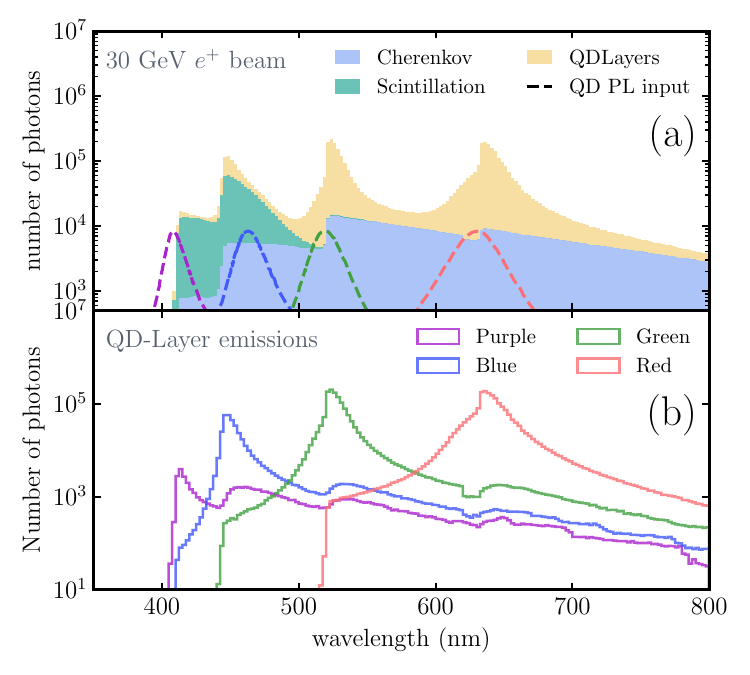}
    \caption{Wavelength distribution of all the photons, generated by $30~GeV$ electron showers, arriving at the rear of the CCAL module (a). Each photon is categorised by source: Cherenkov (blue), scintillation (green), and quantum dot (yellow) processes. Distribution of the photons emitted by each of the QD layers (b).}
    \label{fig:ccal-spectral-reponse}
\end{figure}

The results of our Geant4 simulation demonstrate that chromatic calorimetry can effectively reconstruct shower profiles by capturing the intensities of photons emitted at distinct wavelengths(see Figures.~\ref{fig:ccal-g4-event-display} and \ref{fig:ccal-spectral-reponse}). The clear spectral peaks corresponding to QD emission bands validate the potential of this approach for achieving finer segmentation compared to traditional homogeneous calorimeters. Each peak is selected using the optical filters, and the photons of each peak are counted at the rear of the CCAL module. The response of each of the channels is measured and shown in Figure.~\ref{fig:energy-recontruction}. The figure also shows the energy fraction shared by each module $f_i = R_i/\sum_j R_j$  as a function of the beam energy; it displays a clear progression of the shower profile as the energy increases. Indeed, suppose one observes the energy response of the first channel (red channel). In that case, the energy response decreases, meaning that the shower shares more energy with the subsequent layers, displayed here by a progressive increase of the second layer (green layer) as the beam energy rises. 

\subsubsection{Reconstruction}
The simulations also highlight specific challenges, particularly in the context of light dynamics within the calorimeter. Notably, even when using QDs with minimal reabsorption effects, the refracted light from later stages (bluer layers) tends to be reabsorbed by the earlier crystals. This phenomenon occurs because the QD layers in the calorimeter act as directional pass-through filters for light, allowing light to propagate forward through the layers but not in the reverse direction. We call reflected photons that have been reabsorbed by the previous layers and remitted in a different colour peak, \emph{echo-photons}. For example, in the green layer (second position), about 6\% of the photons emitted by the green layer got reflected at the PWO-PMMA interface, among which 20\% get absorbed by the red layer and re-emitted as a red peak.

Consequently, only 1.3\% of the light emitted by the green layer is from red-echo photons, while similar echo-photon effects are estimated to contribute around 2\% for blue layer and 8\% for purple layer\footnote{It is important to note that PWO emits primarily at $420~\rm{nm}$, resulting in only approximately $\sim 20\%$ overlap between the emission spectrum of PWO and the absorption spectrum of the last QD layer, which peaks at $407 ~\rm{nm}$.}. This characteristic poses a unique challenge for energy reconstruction and estimation, as it impacts how light travels and is captured within the calorimeter, introducing additional nonlinear effects that require further detailed study to optimise the calorimeter's design and functionality. 

\begin{figure}[!ht]
    \includegraphics[width=0.5\textwidth]{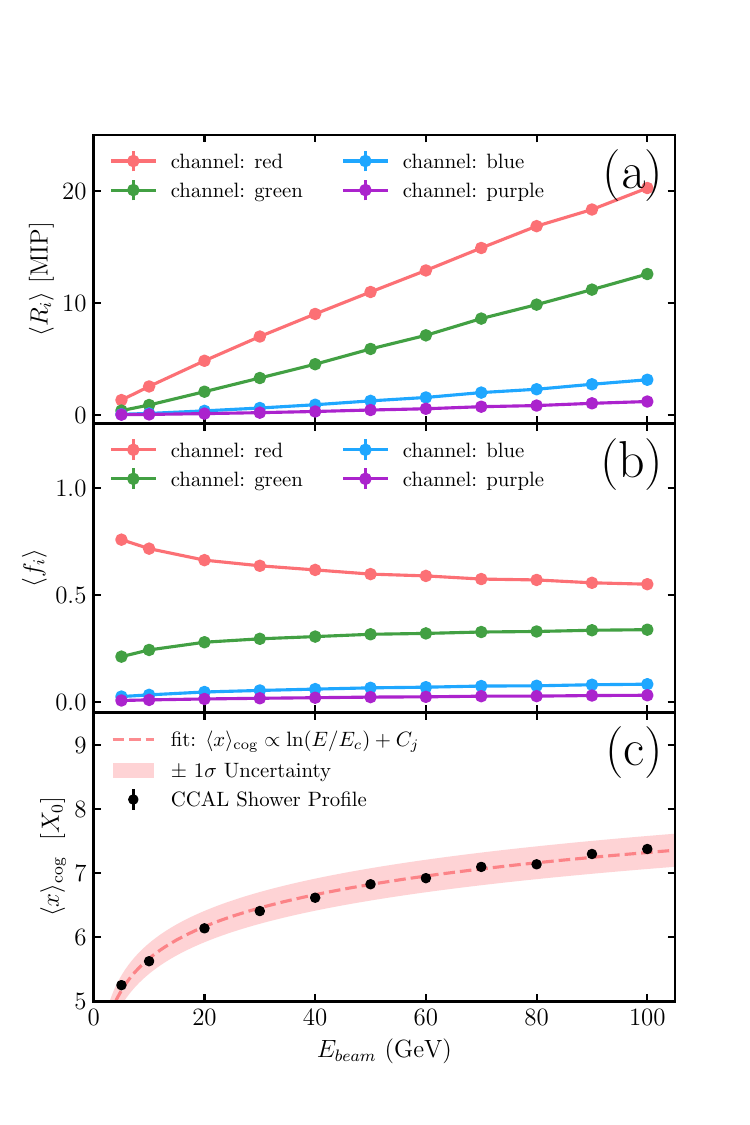}
    \caption{Energy responses $\langle R_i \rangle$ (a) measured as well as the energy fractions $\langle f_i\rangle$ (b) of each of four channels as a function of the beam energy measured in units of MIPs. The bottom plot (c) shows the mean of the longitudinal shower profile $\langle x\rangle_\text{cog}$, or centre of gravity in the longitudinal direction, measured in the unit of interaction length $X_0$. The red line shows a fit that demonstrates the textbook logarithmic dependence of the electromagnetic shower max Ref.~\cite{particle_data_group_review_2022}}
    \label{fig:energy-recontruction}
\end{figure}

Another outcome, shown in Figure~\ref{fig:energy-recontruction}(a), is the energy-dependent variation in spectral channels. At lower energies, light is predominantly from the first "Red" channel; as the beam energy increases, subsequent crystals (Green and Blue) start contributing more to the total reconstructed energy. These trends demonstrate the progression of the shower profile as the beam energy increases. It is helpful to note that these progressions are in units of MIPs, after following the calibration described in section \ref{MIP-Calibration}. 

One can then define the longitudinal shower centre of gravity as $x_\text{cog}\;=\;\sum_{i}f_{i}x_{i}$ where $f_i$ is the reconstructed energy fraction for channel $i$, as defined earlier, centred at depth $x_i$ from the crystal front face for each shower. The quantity, therefore, represents an energy-weighted average depth expressed in radiation-length units $X_0$. The observed rise of $\langle x\rangle_\text{cog}$ with beam energy, as shown in Figure.~\ref{fig:energy-recontruction}, reproduces the well-known expression for the depth of maximum in an electromagnetic shower, $t_\text{max}[X_0]=\ln(E/E_c)+C$ where \textit{E} is the incident electron (or positron) energy, is the PWO critical energy, and $C\approx -0.5$ for electrons (see PDG Sec. 33.2~\cite{particle_data_group_review_2022}). Fitting our simulated points with this functional form (red line in Figure.~\ref{fig:energy-recontruction}) yields a good agreement within statistical uncertainty, confirming that the chromatic-channel reconstruction tracks the longitudinal development predicted by standard shower theory.

\subsubsection{Particle Identification}
The depth-dependent response of the CCAL enables effective differentiation between particle types by leveraging the material properties of PbWO4 and the calorimeter’s design. With a radiation length $X_0 = 0.89 ~\cm$ and interaction length $\lambda_{\text{int}} = 22 ~\cm$, the 27 cm module spans approximately $30 X_0$ and $1.2 \lambda_{\text{int}}$, ensuring full containment of electromagnetic showers and partial containment of hadronic showers. 

\begin{figure}
    \subfloat[][] {\includegraphics[width=0.48\textwidth]{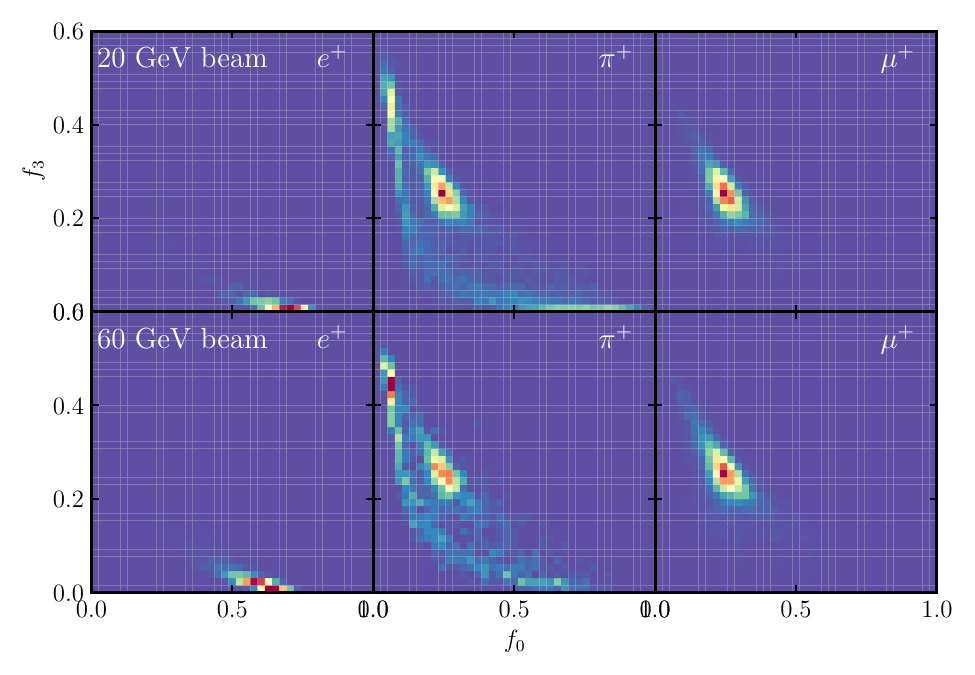}}
    \caption{Amplitude fractions ($f_0$ and $f_3$) for $e^+$, $\pi^+$, and $\mu^+$ at $20~\GeV$ and $60~\GeV$. $e^+$ showers are initiated in the first layer, $\pi^+$ show deeper energy spread, and $\mu^+$ exhibit uniform MIP-like behavior.}
    \label{fig:pid}
\end{figure}

In Figure.~\ref{fig:pid}, the electromagnetic showers initiated by $e^+$ deposit energy primarily in the first few layers ($f_0$), producing compact, early shower profiles. In contrast, hadrons like $\pi^+$ exhibit a dual behaviour: many traverse the module with minimal interaction, appearing as minimum ionising particles, while others initiate deeper hadronic showers, spreading energy across later layers ($f_3$). Figure.~\ref{fig:pid} demonstrates this behavior, showing amplitude fractions $f_0$ and $f_3$ for $e^+$, $\pi^+$, and $\mu^+$ at $20 ~\GeV$ and $60 ~\GeV$. This depth and spectral information highlight the calorimeter’s advanced PID capabilities, which are critical for distinguishing electromagnetic and hadronic particles.

\subsubsection{Energy Measurement and Resolution}

\begin{figure}
    \subfloat[][] {\includegraphics[width=0.25\textwidth]{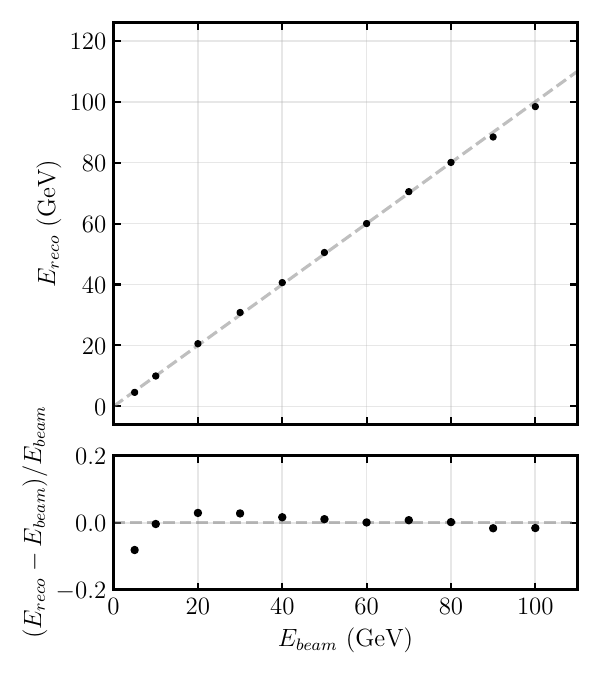}}
    \subfloat[][] {\includegraphics[width=0.25\textwidth]{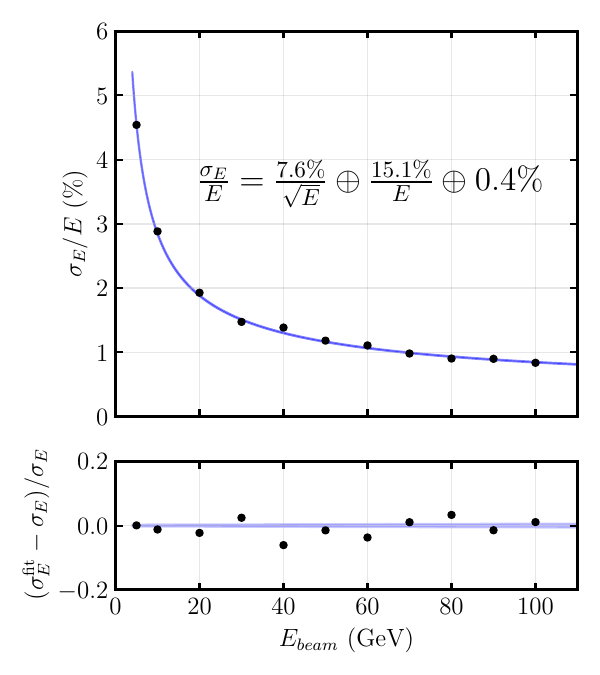}}
    \caption{Chromatic-reconstruction performance of the CCAL prototype for single electrons. (a) Mean reconstructed energy $E_{\text{reco}}$ versus beam energy $E_{\text{beam}}$ in the $5-110 ~\GeV$ range.  The dashed line indicates perfect linearity; the lower panel shows the relative deviation $(E_{\text{reco}}-E_{\text{beam}})/E_{\text{beam}}$, demonstrating that the response is linear to within $5\%$ for $E_{\text{beam}}>10 ~\GeV$ and within $8\%$ at $5 ~\GeV$ when full optical transport is included. (b) Relative energy resolution $\sigma_E/E$ obtained from Gaussian fits to the reconstructed energy $E_\text{reco}$. The blue curve is the standard three-term calorimeter resolution fit, where the terms correspond to stochastic, noise, and constant contributions, respectively.}
    \label{fig:lineariy-resolution}
\end{figure}

A primary goal of a calorimeter is the ability to reconstruct the energy of the incident particle. One should then assess the linearity of the response of the calorimetric channel as a function of the energy of the incident particle. This can be shown in Figure.~\ref{fig:energy-recontruction} (a). The per-channel MIP calibration, described in sub-section\ref{MIP-Calibration}, equalises the response and aims at removing the average echo-photon bias. The event energy is then obtained as a linear combination of the four chromatic amplitudes $E_\text{reco}(\mathbf{w}) = \sum_i^\text{channels}w_i R_i(E_\text{beam})$, with the coefficients $w_i$ chosen to match the energy of the incident particle by minimising $\chi^2 = (E_\text{reco}(\mathbf{w})- E_\text{beam})^2$. Figure~\ref{fig:lineariy-resolution} illustrates the energy response of the CCAL module and its associated energy resolution after the optimisation. With the full optical transport enabled, the response remains linear within  5\% with larger deviations at low energy $\leq 20 ~\GeV$. Since the CCAL’s active QD layers sit at discrete depths, low-energy showers reach maximum after $\sim3 X_0$, so a larger fraction of their
energy is deposited in the first layer. The effective sampling fraction is, therefore, smaller than at higher energies, again pushing the reconstructed energy downwards. The slight downward curvature above $80~\GeV$ stems from shower leakage, which starts being prevalent in these energies. No sensor-saturation model is yet included, so the quoted linearity represents an optimistic limit. Future work will incorporate SiPM pixel depletion and ADC range considerations. The restored linearity allows us to define the energy resolution (constant term of 0.35\%), which agrees with what was measured by the CMS experiment ECAL\cite{noauthor_cms_nodate}.

\subsubsection{Future directions}
While this simulation is encouraging, it has yet to be validated with experimental data. Formulating a unified model of nanophotonic scintillators that encompasses the critical aspects of scintillation processes, such as energy loss by high-energy particles and light emission in nanostructured optical environments, is yet to be developed. An initial effort has already been initiated by Roques-Carmes et al.\cite{roques-carmes_framework_2022}, laying a significant theoretical foundation. However, its practical validation and integration into the Geant4 framework remain key milestones. 

The Geant4 toolkit offers advanced capabilities for simulating particle interactions, but its applications have been limited in some important areas. Specifically, there has yet to be an implementation for simulating quantum dots within bulk scintillators in the context of particle physics, nor for modelling nanophotonic scintillation in many existing experiments. Existing approaches often involve simplifications or generalisations that only partially account for arbitrary types of high-energy particles, scintillator materials, and nanophotonic environments. Therefore, further exploration is needed to bridge this gap. Possible improvements should account for the changes in the confinement energy level with the QD, which will induce either higher or lower emission spectra. Additionally, the treatment of the interaction with ionisation particles is yet to be studied and requires, most probably, dedicated experimental studies. 

In terms of radiation-hardness, A. Erroi et al. \cite{erroi_ultrafast_2023} demonstrated that semiconductor quantum dot–polymer nanocomposite scintillators, specifically CsPbBr$_3$ perovskite nanocrystals embedded in a polyacrylate matrix, can endure extremely high doses of gamma irradiation, up to $1 \rm MGy$, without significant degradation in optical or structural properties. The study demonstrated that the materials maintained stable scintillation performance, photoluminescence efficiency ($\sim90\%$), and structural integrity, exhibiting no signs of damage or performance degradation. This encouraging result suggests that a polymeric QD-doped layer could be sufficient for FCC applications. However, the response to mixed fields (fast neutrons, charged hadrons) is yet to be validated. A dedicated experimental programme to characterise both the polymer matrix and the embedded quantum dots under representative radiation spectra is planned. Quantitative radiation-hardness simulations are therefore deliberately omitted here, as the present paper focuses on demonstrating the chromatic-reconstruction technique rather than on materials qualification.
\newline
\newline 
\section{Conclusions}

In this paper, we have introduced the concept of chromatic calorimetry, which utilises the unique optical properties of quantum dots to address the limitations in timing resolution and longitudinal segmentation inherent in current calorimetric technologies. Our design proposal explored two different approaches: direct embedding of quantum dots in scintillator crystals and a hybrid approach combining QD-doped organic polymers with high-Z scintillators. Detailed simulations using the Geant4 toolkit demonstrated the feasibility of this technology, showing clear spectral peaks corresponding to the QD emissions and promising enhancements in electromagnetic shower profile reconstruction. Despite these promising results, challenges such as photon reabsorption and the lack of existing models for QD interactions in Geant4 highlight the need for further research. 

Moving forward, experimental validation of our simulation model and the development of new nanophotonic scintillator frameworks will be crucial in bringing chromatic calorimetry to the forefront of high-energy physics experiments. The results presented in this work provide a significant step towards realising more precise and efficient calorimetric detectors, paving the way for future innovations in particle detection and analysis at next-generation colliders.



\bibliography{bibliography}
\bibliographystyle{cms_unsrt}
\end{document}